# Bridging User Feedback and System Diagnosis: Reproducing Mobile Performance Issues from Reviews

ZHENGQUAN LI, The Hong Kong University of Science and Technology (Guangzhou), China
ZHENHAO LI, York University, Canada
SIDONG FENG, The Chinese University of Hong Kong, Shenzhen, China
CUIYUN GAO, Harbin Institute of Technology, China
TAO ZHANG, Macau University of Science and Technology, China
ZISHUO DING*, The Hong Kong University of Science and Technology (Guangzhou), China

Mobile application performance is a vital factor for user experience. Yet, performance issues are notoriously difficult to detect in development environments, where they often manifest less conspicuously, making their diagnosis more challenging. In this setting, app reviews from end users across diverse devices and usage contexts can provide timely and context-rich information about emerging performance issues. However, unlike structured bug reports, app reviews are written by end-users and tend to be more ambiguous, with individual reviews often providing only partial descriptions of the underlying issue. To bridge this gap, we present **RevPerf**, the first approach to automatically reproduce mobile application performance issues by leveraging and synthesizing information from app reviews. **RevPerf** retrieves complementary reviews via semantic retrieval and uses prompt engineering to integrate them, enriching the original review with performance issue details. An execution agent is then employed to generate and execute commands to reproduce the issue. After executing all necessary steps, the system incorporates multifaceted detection methods to identify performance issues by monitoring Android logs, GUI changes, and system resource utilization during the reproduction process. Experimental results demonstrate that our proposed framework achieves a 72.73 % success rate in reproducing performance issues on the constructed dataset, outperforming the best baseline by 27.28 percentage points.

## 1 Introduction

Mobile applications have become integral to daily life, with Android alone serving over three billion users globally [18]. Application performance is critical to user experience [38, 59], as performance

*Corresponding author.

Authors' Contact Information: Zhengquan Li, The Hong Kong University of Science and Technology (Guangzhou), Guangzhou, China, zli169@connect.hkust-gz.edu.cn; Zhenhao Li, York University, Toronto, Canada, lzhenhao@yorku.ca; Sidong Feng, The Chinese University of Hong Kong, Shenzhen, Shenzhen, China, sidongfeng@cuhk.edu.cn; Cuiyun Gao, Harbin Institute of Technology, Shenzhen, China, gaocuiyun@hit.edu.cn; Tao Zhang, Macau University of Science and Technology, Macau, China, tazhang@must.edu.mo; Zishuo Ding, The Hong Kong University of Science and Technology (Guangzhou), Guangzhou, China, zishuoding@hkust-gz.edu.cn.

issues such as responsiveness delays [19, 31, 39], excessive battery drain [13, 15], or redundant memory usage [24, 53] can severely degrade user satisfaction [23, 44, 45].

While performance issues in Android apps have received significant attention in current software engineering research [43, 59], the reproduction of performance issues remains a challenge due to multiple reasons. Performance issues often exhibit intermittent manifestation and stem from complex factors [38, 42], compounded by Android's fragmentation across different environments [18]. Many issues only emerge under specific operational scenarios [14, 60], making stable reproduction in laboratory settings exceedingly difficult. Due to constraints on testing resources and time, developers cannot cover all potential performance issues during development phases. Collectively, these factors make manual performance debugging and issue reproduction arduous, requiring substantial human expertise and time [38].

In this context, the vast corpus of user reviews accumulated in app stores provides a vital way for developers to detect performance issues. These reviews provide direct, in-the-field feedback from a broad base of end users encountering real-world problems across diverse devices and usage contexts. Despite their noisy and unstructured nature, they often reveal early signals of user-perceived performance issues [48, 56]. A real-world example from the Markor [9] app is illustrated in Fig. 1, where a user complains that file reading times are significantly longer compared to other applications. When testing a 3 MB file in Markor and examining the corresponding Android system logs, we observed that activity creation, which encompasses the file reading process, takes over 3 seconds. This finding in the logs confirms the user's complaint and provides objective evidence of performance degradation. Analyzing user reviews has become a common approach for bug reproduction [22, 47]. Early work like RepRev [34] used natural language processing (NLP) to extract information from reviews for crash bug reproduction. Recent advances have incorporated LLMs into this process, with tools like AdbGPT [21], ReBL [57], and ReActDroid [27] demonstrating the potential of LLM-based bug reproduction frameworks.

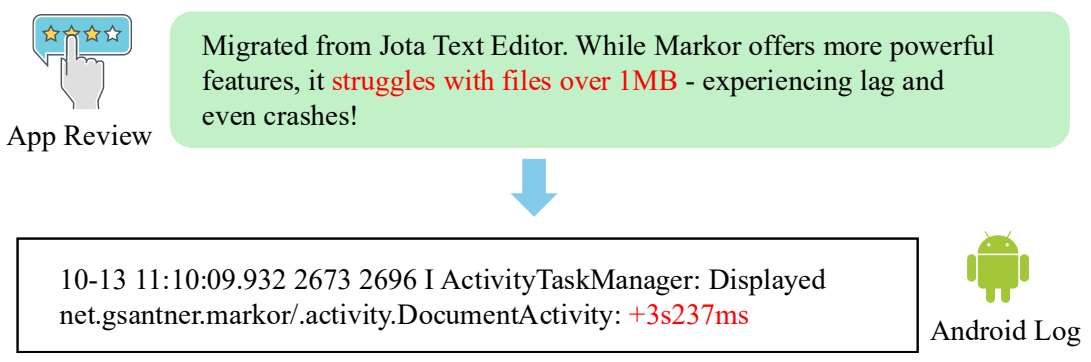

Fig. 1. Performance issue in Markor Editor: user complaint and validating system logs.

However, existing automated bug reproduction frameworks primarily focus on crash issues [63, 64] and functional bugs [16, 57], neglecting performance issues. This oversight stems from three main limitations. First, existing automated bug reproduction datasets [54, 62–64] predominantly focus on crash and functional bugs, leaving a scarcity of datasets for performance issues. Second, existing performance issue detection methods [33, 43, 51] rely on category-specific metrics, such as frame latency for lag [43] and power anomalies for energy issues [36], and thus generalize poorly across diverse performance issues. This fragmentation also complicates LLM-based automation, which must navigate multiple specialized tools and interpret heterogeneous diagnostic outputs.

Motivated by these gaps, we conduct manual analysis on our collected dataset to identify three key challenges (see Section 2.3 for details) in leveraging LLMs to reproduce and detect performance issues. **The first challenge is the lack of structured and sufficient context in user reviews.** User reviews are written by non-technical users who do not follow any structured format and often include noisy content such as emotional expressions, praise, or complaints unrelated to the issue itself. Moreover, unlike formal bug reports, user reviews often omit critical technical details such as the specific UI screens involved. **The second challenge involves hidden environmental and workload dependencies.** Performance issues can depend on specific runtime conditions that remain implicit in reviews. These include complex user interactions (e.g., rapid typing, repeated rotations) as well as system-level and in-app configurations. Successful reproduction requires configuring these settings appropriately and executing the corresponding actions correctly before performing the reproduction steps. **The third challenge arises from the inherent ambiguity and subjectivity of performance complaints.** Users describe issues using imprecise, non-technical terms like "slow" or "laggy", which cannot be directly mapped to specific system metrics. Moreover, unlike crashes that produce clear failure signals, performance degradation lacks objective thresholds and manifests through diverse symptoms (e.g., UI lag, frame drops, unresponsiveness).

To address the above limitations, we propose **RevPerf** (Review-based Automated Performance Bug Reproduction), a novel framework for automatically reproducing and detecting performance issues from user reviews. To our knowledge, this is the first framework for automatically reproducing and detecting performance issues from user reviews. **RevPerf** employs a semantic-based retrieval method to gather relevant information from reviews of the same app version to enrich the original review. It then utilizes an execution agent to generate and execute commands in the Android emulator to reproduce performance issues and collect necessary information, including the log messages and system resource consumption over time. Finally, during the detection phase, it automatically combines the reports from different detection modules and the execution history to determine whether the performance issue has been triggered. Subsequently, it generates an analysis report to conclude the reproduction, reporting whether the expected performance issues were triggered.

To evaluate our framework, we construct a dataset of 95,067 app reviews, including 616 high-quality reviews validated by developer responses or GitHub issues. Among these high-quality reviews, 55 were manually verified to contain reproducible performance issues by the authors. We then conducted a series of experiments by applying **RevPerf** to this dataset to assess its performance across multiple dimensions. In the results, **RevPerf** successfully reproduces 40 of the 55 performance issues, achieving a 72.73 % success rate. The average reproduction time was 146.66 seconds, demonstrating the framework's efficiency in extracting performance issues and executing reproduction steps. Furthermore, we conducted an ablation study to evaluate the contribution of each component. The experimental results demonstrate the robustness and effectiveness of our framework for reproducing performance issues.

The contributions of this paper are as follows:

- We conduct an empirical study on the collected app reviews and identify several key challenges in reproducing performance issues from them. Based on this analysis, we construct a dataset of 55 reviews with manually reproduced performance issues. This dataset is available to support future research [37].
- We propose **RevPerf**, the first framework to automatically reproduce performance issues from app reviews. **RevPerf** employs a semantic-based retrieval method to enrich the information of the target user review, and three detection approaches to detect different types of performance issues.
- We conduct extensive experiments to evaluate our framework on the constructed dataset. **RevPerf** achieves a 72.73 % success rate (40/55), outperforming the best baseline by 27.28 percentage points, with an average reproduction time of 146.66 seconds.

## 2 Preliminary Study of Performance Issue Reproduction from Reviews

Given the limited research [59] on leveraging app reviews to reproduce performance issues, we conduct a preliminary study to better understand the characteristics of such issues in mobile applications and the challenges of reproducing them from user feedback.

### 2.1 Data Collection

*2.1.1 Studied Apps.* We start by collecting apps referenced in recent studies on automated bug reproduction [21, 27, 57, 58] and widely used in crash or functional bug reproduction datasets [54, 63, 64]. We then select apps based on two criteria: (i) they are actively maintained on Google Play and have at least 1,000 user reviews, and (ii) they are open-source on GitHub, with accessible historical versions and runnable builds on Android emulators.

This process yields a set of 29 open-source apps spanning multiple categories, including *Markor* (note-taking) [9], *AnkiDroid* (flashcard learning) [2], and others.

*2.1.2 Developer-Attended Reviews.* Unlike prior work [34] that treats all user reviews as a monolithic source of feedback, we focus on a curated subset: *developer-attended reviews*. These are reviews that have either received a developer reply on Google Play or have been referenced in GitHub issues. Such reviews highlight performance problems that are not only visible to users but also recognized by developers as important and actionable. Specifically, we first collect all user reviews using the *google-play-scraper* library [7], resulting in 95,067 reviews from the previously studied 29 apps. We then filter the reviews to retain only those related to performance issues, using a list of 43 keywords derived from prior work [45, 59] and expanded through manual inspection. Specifically, we first extracted performance-related terms from prior studies, covering common symptoms such as slow response, lag, freezing, unresponsiveness, and excessive resource consumption. We then normalized semantically equivalent expressions and removed duplicates to form an initial keyword set. To better capture user-facing descriptions in app reviews, we further inspected a random sample of collected reviews and added commonly used lay expressions that were not explicitly covered by prior work. This keyword-based filtering is used only as a high-recall first-stage filter before the subsequent LLM-based classification and manual verification. Reviews containing at least one keyword are selected, reducing the dataset to 5,661 entries.

While keyword filtering helps reduce irrelevant content, it may still yield false positives [34, 59]. To improve accuracy, we employ Llama3.3:70b [55] to further filter reviews that describe performance issues. The prompt assigns the LLM the role of a performance-issue analysis expert and provides performance-related cues based on our keyword construction process, including load time, freezing, memory usage, and other symptoms related to resource consumption or responsiveness. The model is then asked to determine whether each review contains performance-related content and to output a binary decision with a brief rationale. We use Llama3.3:70b because it is a widely used open-weight LLM suitable for semantic filtering over noisy and informal user reviews, which also improves the reproducibility of our dataset construction. This step results in 3,094 performance-related reviews. To assess the reliability of this LLM-based filtering step, we randomly sampled three non-overlapping sets of 200 reviews from the 3,094 filtered reviews, yielding 600 sampled reviews in total. Two authors independently annotated the sampled reviews to determine whether they indeed describe performance-related issues. The two authors confirmed that 98.67 % and 98.17 % of the sampled reviews are true performance-related reviews, respectively, with an overall agreement of 99.50 % (597/600) and a Cohen's $\kappa$ of 0.84. The three disagreements were resolved through discussion between the two authors to produce the final labels. These results suggest that the LLM-based filtering step reliably identifies performance-related reviews.

Finally, the dataset consists of 623 reviews from 29 apps, including 576 with developer replies and 47 referenced in GitHub issues. These reviews reflect developer-recognized cases with higher potential for reproduction.

### 2.2 Manual Characterization of Performance Issue Reviews

Based on the collected developer-attended reviews, we conduct a manual examination to bridge the gap between unstructured descriptions and a systematic categorization that can support both diagnosis and reproduction of performance issues.

*2.2.1 Methodology of Manual Analysis.* Similar to prior work [59], two authors are involved as annotators, and our manual study process consists of three steps:

**Step 1:** The first annotator reviews 623 reviews and notes key elements such as performance symptoms (e.g., lag) and proposes an initial set of categories to describe performance issues grounded in a performance taxonomy established by prior research [32, 38, 45]. Note that during this process, seven reviews initially selected are excluded, as they either express feature requests or report problems related to other software. This filtering results in a final dataset of 616 reviews.

**Step 2:** A second annotator reviews the initially proposed categories. The category refinement process is informed by prior studies [40, 59] on app reviews and considers factors relevant to issue reproduction, such as specific triggers, or observable conditions.

**Step 3:** The two annotators independently labeled each review using the finalized categories. To assess the reliability of the independent category annotation, we computed Cohen's $\kappa$ before disagreement resolution, which yielded a value of 0.7581. The annotators then discussed all disagreements and reached consensus to determine the final category labels.

*2.2.2 A Categorization of Performance Issues.* As a result, three principal categories are distinguished on the basis of their modes of reproduction:

**Slow Interaction and Delayed Response.** This category includes UI animation lag, slow loading or processing time, and delays in functional response that negatively affect user experience. These issues typically do not block functionality but cause noticeable delays that exceed user expectations. Users often identify such problems by comparing them to previous usage patterns, earlier application versions (i.e., performance regressions), or similar applications (cf. Fig. 1). For example, a user from the messaging application QKSMS reported "App was great until my phone updated it yesterday automatically. Now when I type enough to go to a new line, it doesn't automatically scroll down until I pause typing for a second, so until I manually stop, I'm typing blind. ..." These issues may happen when operations that require heavy computation or processing occur on the main UI thread (the target thread responsible for updating the interface). While such UI lags stem from diverse root causes, they often yield consistent system log patterns, as the Android framework records most rendering disruptions through standardized alerts (e.g., Choreographer frame-drop warnings). In terms of loading and processing issues, system logs offer precise timing information regarding the execution durations of various tasks, as demonstrated in Fig. 1.

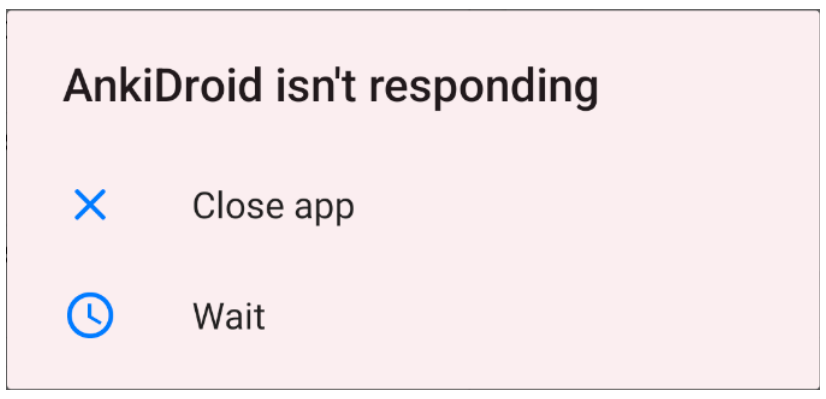

Fig. 2. An example of the ANR error in AnkiDroid app.

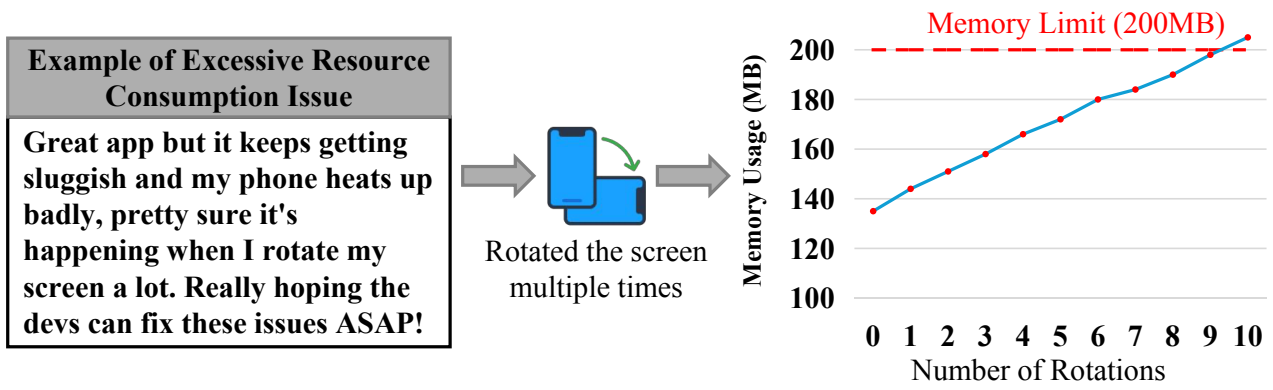

Fig. 3. A review reports a memory leak issue.

**App Freezing and Unresponsiveness.** This category covers more severe issues that impair app stability or even functionality. It includes Application Not Responding (ANR) events, interface freezing, and functional failures. Unlike the responsiveness delays in the previous category, these issues render the app or its specific component completely unresponsive, often for more than 5 seconds [3]. For instance, a user reported a usability issue in an app review for the podcast application AntennaPod, stating, "During playback pause play button stops working after screen off and on." This bug prevents the user from managing playback directly from the main interface.

Fig. 2 also shows an ANR event in the AnkiDroid [2] app, where the app becomes unresponsive to user input. We classify such critical degradation under this category.

**Excessive Resource Consumption.** This category captures performance issues related to inefficient use of device resources, such as CPU and memory. High CPU usage may indicate unnecessary or inefficient computation, while increased memory usage can suggest poor memory management (e.g., memory leaks). Battery drain and network data overuse are also reported by users in reviews. A real example from SMS app QKSMS [10] is shown in Fig. 3, memory usage may rise continuously in memory leak scenarios, especially when certain operations are repeated, indicating improper resource cleanup. When the heap memory usage exceeds the predefined limit in the emulator, the app will be terminated by the Android system. To identify such anomalies, it is essential to track resource fluctuations throughout the reproduction sequence. By correlating each reproduction step with its corresponding resource footprint, one can determine whether specific operations lead to abnormal consumption patterns.

Fig. 4 illustrates the distribution of performance issue categories across the 616 developer-attended reviews. Among them, App Freezing and Unresponsiveness issues are the most commonly reported, followed by Slow Interaction and Delayed Response issues, and Excessive Resource Consumption issues. Notably, a substantial number of reviews fall into multiple categories, indicating that many performance issues are multi-dimensional. For example, 141 reviews report problems related to both slowness and unresponsiveness, while 9 reviews are labeled with all three categories. This indicates that a single code issue can lead to multi-dimensional performance problems, spanning multiple categories. This overlap highlights the complexity of user-perceived performance issues and suggests that effective reproduction and diagnosis often require reasoning across multiple performance aspects.

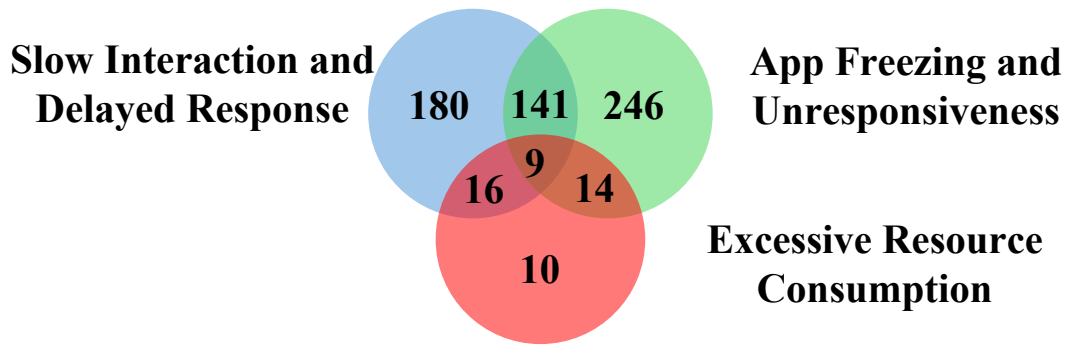

Fig. 4. Overlap among the three performance issue categories in the 616 reviews.

## 2.3 Challenges and Implications for Reproducing Performance Issues

In this section, we summarize the key challenges identified through our manual analysis and discuss their implications for automated reproduction and detection. These challenges arise not only from the content and quality of the user reviews themselves but also from the inherent nature of performance problems, which are often environment-dependent, vaguely described, and multi-dimensional.

**Challenge 1: Lack of Structured and Sufficient Context.** User reviews are often noisy, brief, and high-level, lacking the structured information needed to support reproduction [34, 59]. Unlike formal issue reports, which may include step-by-step instructions, affected components, or expected behavior, user reviews tend to describe symptoms in vague terms. This is largely due to users' limited technical background and the informal nature of app store feedback. For example, as shown in Fig. 1, a user reports that the Markor app "struggles with files over 1MB – experiencing lag and even crashes", but omits key contextual information such as the UI screen involved, the actions taken prior to the lag, or the exact reproduction conditions. Without such details, it is difficult to recreate the issue reliably.

**Implication:** Effective reproduction frameworks cannot rely on a single, isolated review. They must incorporate mechanisms to infer or enrich missing context. This challenge also calls for advanced natural language processing techniques to extract and interpret reproduction-relevant information from noisy and incomplete user feedback.

**Challenge 2: Hidden and Complex Environmental and Workload Dependencies.** While examining the app reviews, we also observe that the performance issues reported in user reviews are dependent on specific runtime conditions, such as device settings, network quality, or user workload. These dependencies often remain implicit or vaguely described and span a range of factors: (i) **complex user interactions**, such as rapid typing or repeated screen rotation, which are challenging to model or automate; and (ii) **system and application settings**, such as system-level accessibility options (e.g., the "Remove animations" toggle) and in-app UI theme configurations (e.g., light and dark mode), which can significantly alter an application's runtime behavior.

**Implication:** Reproduction frameworks must consider both observable and implicit contextual factors. This involves checking system conditions before execution and simulating realistic workloads, rather than merely replaying static user actions.

**Challenge 3: Ambiguity and Subjectivity in Performance Complaints.** Unlike crash bugs that produce clear and deterministic failure signals [21, 57], performance issues are often ambiguous and highly subjective [45]. This ambiguity appears in three main forms:

(1) **Subjective user language:** Our analysis shows that users tend to describe problems and actions using non-technical terms such as "slow", "hang", or "hot" (e.g., overheating or battery drain). These terms lack specificity and introduce a semantic gap that makes it difficult to directly map the descriptions to concrete system metrics.
(2) **Absence of a clear failure signal:** Performance issues in reviews typically do not produce binary indicators. Instead, they are characterized by degradation relative to user expectations, such as slower behavior compared to previous usage patterns, earlier versions, or peer applications. Meanwhile, a single symptom is often insufficient to confirm the presence of a performance issue. For instance, a delayed response time might be caused by expected computation demands rather than an underlying performance problem.
(3) **Multi-faceted manifestations:** Performance problems can manifest through a wide variety of symptoms, such as UI lag, frame drops, or unresponsiveness. These symptoms differ in severity, visibility, and root cause, making it difficult for any single detection strategy to capture them all. Furthermore, a single underlying performance issue may exhibit multiple concurrent symptoms. Our analysis found that 29.22 % (180/616) of the reviews mentioned multiple types of issues within a single report. This multiplicity increases the difficulty of detection, as tools limited to identifying a single performance signal may overlook the broader problem.

**Implication:** To handle the inherent ambiguity of performance complaints [59], reproduction tools must adopt diverse detection strategies and establish performance baselines. Relying on a single fixed error pattern is insufficient. Instead, frameworks should interpret subjective user feedback, monitor multiple performance dimensions, and correlate varied symptoms to determine whether an issue truly exists.

## 3 Approach

To address the identified challenges in reproducing performance issues from user reviews, we propose **RevPerf** that automatically analyzes, reproduces, and diagnoses performance problems in Android applications using user feedback.

### 3.1 Approach Overview

As shown in Fig. 5, **RevPerf** consists of three main stages: Review Augmentation, Feedback-Driven Execution, and Performance Issue Detection. Given a user review, **RevPerf** retrieves related reviews and app descriptions to provide context. It then generates and executes a sequence of commands in an Android emulator, refining actions based on emulator feedback to improve reproduction success. During the Issue Detection stage, **RevPerf** collects runtime data (e.g., log messages, system resource, UI state) and uses an Issue Detector to identify performance issues. If successful, it outputs a structured report detailing the issue. Otherwise, it returns a diagnostic explanation for further refinement or manual inspection.

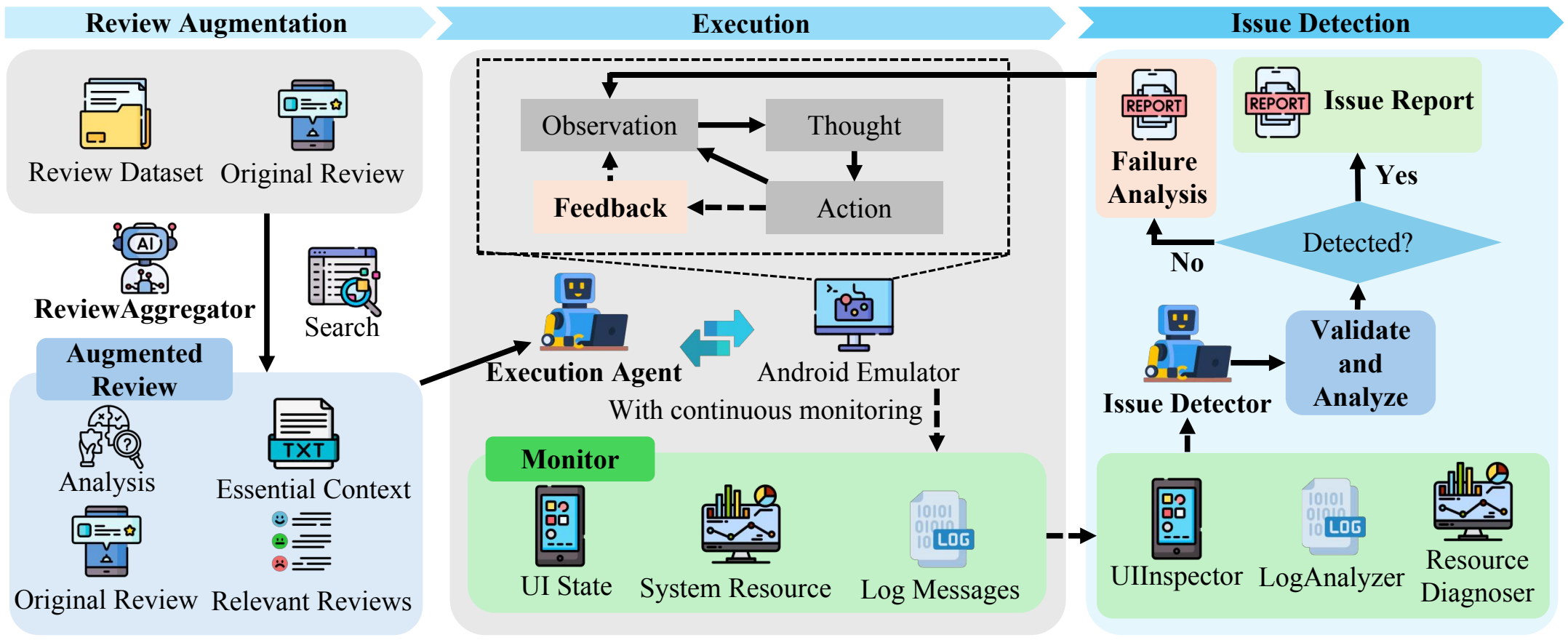

Fig. 5. Overview of **RevPerf**.

### 3.2 Review Augmentation

Although user reviews often contain noise and informal language, prior studies [34, 59] have shown that they still carry valuable information. Therefore, to address the **lack of structured and sufficient context (Challenge 1)**, we design a review augmentation module, ReviewAggregator, that enriches the original review with contextual information. The goal is to help the LLM better understand the performance issue by incorporating similar reports and app-specific knowledge.

The first step is a semantics-based retrieval process, where the original review is encoded into a vector using a pre-trained embedding model. We then identify the top-$k$ most similar reviews based on cosine similarity from the same app version as potential candidates. Note that we limit the retrieval to the same version because such reviews are more likely to describe the same underlying issue and share a consistent execution context, and thus avoiding inconsistencies caused by feature changes or UI updates.

In the second step, we further filter these candidate samples to reduce potential noise, as previous studies have demonstrated that irrelevant or noisy context can negatively impact the performance of reproduction [25, 65]. Specifically, we use an LLM-based relevance filter. The prompt takes the app name, the original review, and the retrieved candidate reviews as input, assigns the LLM the role of a performance-issue expert, and asks it to retain only reviews describing the same or highly similar performance symptom, trigger condition, affected feature, or complementary reproduction cue. To guide this judgment, we adopt category-guided prompting by listing representative performance-issue symptoms discussed in Section 2.2.2 as relevance criteria, helping the LLM focus on reproduction-relevant semantic similarity rather than superficial lexical similarity. The retained reviews are then aggregated with the original review, app name, and software description to generate the enriched review, which contains two parts: an analysis of the reported performance issue characteristics and a high-level behavioral objective for UI exploration. The latter guides the subsequent Execution Agent toward the target interface and performance state without imposing a concrete action sequence.

### 3.3 Feedback-Driven Execution

In this stage, we employ an execution agent based on the iterative "Observation–Thought–Action" paradigm [27, 61] to simulate user interactions within an Android emulator. While this paradigm is well-established in the literature, our primary contribution lies in three key enhancements aimed at addressing **Challenge 2** (complex environmental and workload dependencies) and **Challenge 3** (ambiguous and subjective performance complaints): (i) an expanded set of actions to simulate complex user interactions, (ii) mechanisms to interpret and enforce environmental constraints from the review, and (iii) a runtime tracing component for continuous system monitoring.

*3.3.1 Observation.* In the Observation phase, the agent collects various contextual information, including the current GUI hierarchy (in XML format), the enriched review from the previous Review Augmentation stage, and a list of available operations. In subsequent iterations, additional context, such as past actions, emulator feedback, and current task status, is incorporated to assist the agent in adjusting the execution plan and determining the next command to execute.

*3.3.2 Thought.* The Thought phase is the primary decision-making stage. In this phase, the agent interprets observations and generates the next command in textual form, which is then passed to the Action phase for execution. In each iteration, the agent plans its next action based on the review-reported trigger action (and its UI context), the execution history, and the UI states explored so far. The review-reported trigger action and its UI context are inserted into the execution agent's initial prompt as the key information for reproduction. For the execution history and the UI states explored so far, we maintain an agent–emulator interaction log that accumulates all past actions and outcomes and is appended to the prompt at each iteration as context for the current reasoning step.

The same prompt template also explicitly includes guidelines that instruct the agent to consult the original review for any required settings, including system-level and in-app settings, before navigating to the target screen to perform the trigger action. If the review specifies such settings, the agent applies them first. Otherwise, it proceeds directly with planning and reproducing the issue. Once the agent reaches the target screen, its goal shifts to plan the review-reported action sequence (e.g., entering a value in a text field and tapping the corresponding button) and generate the corresponding commands. Following previous study [57], the agent further leverages execution feedback to adapt its strategy: if an action fails to execute as expected, it explores alternative UI elements or backtracks to previously visited UI states. Failed attempts are logged to avoid redundant actions.

Additionally, the agent monitors for two key conditions to transition to the Issue Detection Phase: (1) when the current GUI state matches the performance issue description from the review, indicating potential issue manifestation, or (2) when all reproduction steps have been successfully executed, requiring verification of whether the performance issue has been triggered.

*3.3.3 Action.* In the Action phase, the agent parses and validates the generated command from the Thought phase, then executes it in the emulator using a set of predefined operations. The execution status of each command (e.g., "Failed to execute <command_name>" or "Successfully executed <command_name>") is returned to the Observation stage as system feedback, helping the agent adjust subsequent actions. Following previous frameworks [27, 57], we implemented ten basic actions, including `restart, scroll, rotate, back, swipe, multiple_selection, navigate_up, click, long_click, set_text.` Note that to generate commands that support complex interaction scenarios, we extend the action set with additional commands such as `wait`, `input_long_text`, `open_app_drawer`, `lock_screen`, and `unlock_screen`, enabling it to handle behaviors involving delayed responses, cross-app transitions, and large-scale text input. The

`input_long_text` instruction inputs extremely long text sequences (2,000,000 characters) to expose potential performance bottlenecks in applications with inefficient text processing, referring to the relevant documentation [5]. Moreover, `open_app_drawer` allows the agent to switch from the foreground app to the App Drawer for cross-app switching. Notably, **RevPerf** records changes in system resource usage, log messages, and UI states throughout execution. We describe these monitoring signals in detail in Section 3.4.1.

### 3.4 Issue Detection

Once the execution process is complete, **RevPerf** transitions to the Issue Detection stage, where it determines whether a performance issue has occurred based on the runtime data collected during execution. The central task here is to address **Challenge 3**: ambiguity and subjectivity in performance complaints, which distinguishes this work from crash-focused reproduction that relies on explicit failure signals. To tackle such challenges, we design an **Issue Detector** that integrates three complementary analyzers: a LogAnalyzer for analyzing log messages, a ResourceDiagnoser for evaluating runtime metrics, and a UIInspector for examining visual state changes. This multifaceted approach enables flexible and reliable diagnosis by correlating varied and often subtle symptoms, moving beyond the limitations of fixed error patterns. Specifically, we integrate the execution agent with an ADB-based monitoring mechanism that continuously collects runtime signals, including log messages, UI state transitions, and system metrics (e.g., heap usage, CPU utilization, memory usage, swap usage, and the number of active activities). These data streams provide a comprehensive view of the app's runtime behavior and are forwarded to the Issue Detector for analysis. This design treats Issue Detection as post-execution diagnosis rather than online monitoring. During execution, **RevPerf** focuses on lightweight evidence capture, while LLM-based reasoning is performed only after the trace has been preserved. As a result, LLM inference latency may affect when **RevPerf** reports a diagnosis, but it does not affect whether short-lived runtime symptoms are recorded for analysis.

*3.4.1 Design of the Integrated Analyzers.* Below, we detail how each analyzer targets a distinct class of performance symptoms and contributes complementary evidence to support the detection process:

(i) **LogAnalyzer** identifies anomalous entries and system warnings that are indicative of performance issues. **RevPerf** captures Android log messages using Logcat [8] to obtain system- and application-level information during the action stage. These logs include developer-inserted messages, activity lifecycle events, and function-level execution details. For mobile applications, logs are particularly valuable as they provide a chronological record of runtime events that capture both system-level and app-specific behaviors. Since performance issues often manifest as subtle deviations from normal usage patterns rather than explicit failures (Challenge 3), this sequential data is essential for robust issue detection. Inspired by the widespread use of log-based anomaly detection, our LogAnalyzer leverages an LLM to interpret log context and detect abnormal sequences. This enables it to uncover critical symptoms, such as skipped frame warnings, rendering delays, or ANRs, that might otherwise be overlooked. Fig. 6 shows a real example from a two-factor authentication app Aegis, where the LogAnalyzer successfully correlates log entries with lag frames observed during runtime.

(ii) **ResourceDiagnoser** analyzes runtime metrics (e.g., heap usage, CPU utilization, and live Activity instance count) to detect anomalous behaviors over time. Specifically, it records per-action resource usage to form temporal sequences of metrics, and leverages an LLM to evaluate whether these sequences exhibit sustained growth that may indicate issues such as memory leaks. More generally, this approach enables ResourceDiagnoser to surface a broad range of abnormal trends beyond memory leaks (e.g., an activity count that keeps increasing under repeated back-and-forth operations and atypical memory-related indicators reported by `dumpsys meminfo`).

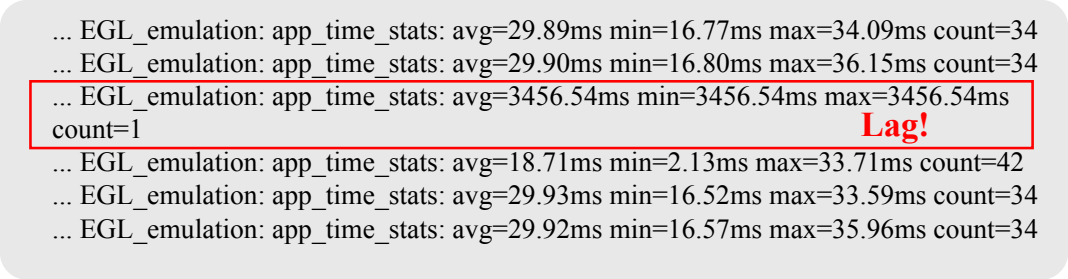

Fig. 6. An example of how lag is identified in log entries.

(iii) **UIInspector** analyzes the evolution of UI states during execution. It compares UI hierarchy before and after each action to detect visual symptoms such as UI stalling, unresponsive interfaces, or abnormal state transitions. This component is particularly useful for capturing issues that are directly observable from the GUI, complementing log- and resource-based evidence.

*3.4.2 Design of Detection Prompts.* Each analyzer employs a standardized template that incorporates distinct background knowledge and instructions to guide the LLM in identifying and analyzing issues. As shown in Fig. 7, every prompt consists of five sections: Task Overview, Background Knowledge, Chain of Thought Prompting, Examples of Successful and Failed Outputs, and Enriched Review. The complete prompts used by all analyzers are available in our open-source artifact [37]. This prompt structure enables the LLM to understand the task and produce a formatted output for subsequent parsing. We begin with a concise task description to provide the LLM with an initial understanding. Next, by incorporating unique background knowledge for different analyzers, we guide the agent in recognizing performance issues in the logs, UI changes, and system resource changes. LogAnalyzer's template includes background knowledge related to the logs. For the ResourceDiagnoser, the prompt incorporates an introduction to the meaning of each metric derived from ADB commands. The background knowledge in the prompt of the UIInspector contains GUI analysis prerequisites. Additionally, the template includes examples of successful and failed output formats. The parsing function processes these formatted outputs to extract detection results, including performance issue trigger status, specific error types, and complete analysis reports. The Issue Detector validates these reports to generate the final assessment.

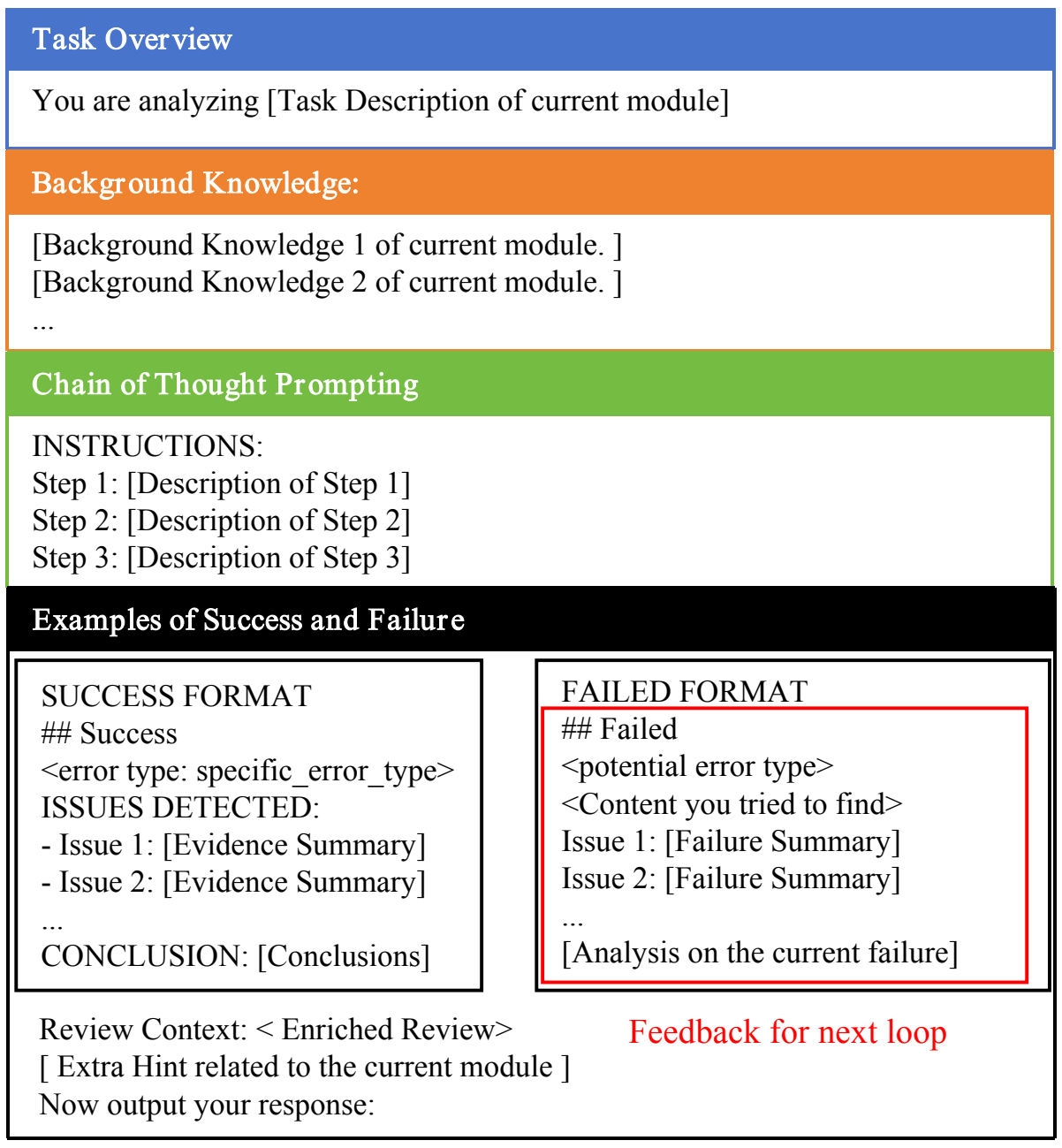

Fig. 7. The standardized prompt template used by the Issue Detector across LogAnalyzer, ResourceDiagnoser, and UIInspector.

When a performance issue is successfully detected and corresponding evidence is found, the Issue Detector outputs a comprehensive reproduction report. This report includes the type of performance issue, the total number of operations executed during reproduction, the total duration of the reproduction process, the evidence confirming successful reproduction, and the associated analysis process. Such evidence may include a set of resource utilization records pertinent to the performance issue, relevant rendering logs, or observed changes in the GUI state (e.g., before and after the presumed trigger), among other indicators. After three consecutive failed reports from all analyzers, the framework terminates the reproduction process and generates a failure analysis report to reduce inference costs and prevent indefinite execution.

## 4 Experiment

### 4.1 Experiment Dataset

Based on our collected dataset of 616 high-quality app reviews, we construct a set of performance issues that can be successfully reproduced in an emulator environment. However, not all review-described issues can be reproduced reliably. For several apps (e.g., Vespucci, WikiMed, and Open Food Facts), the corresponding historical versions are no longer obtainable. For others (e.g., Nextcloud and WordPress), many performance issues depend on specific user data and interactions with remote server-side services. We therefore manually validate each candidate by downloading the APK version mentioned in the review and reproducing the reported behavior in an emulator. This process yields 55 reproducible cases, which we use to evaluate **RevPerf**.

### 4.2 Implementation

All experiments were performed on a MacBook Air featuring an Apple M1 chip, 16 GB of RAM, and 256 GB of storage, running macOS version 15.4.1. We utilized Gemini-2.5-Flash as the core LLM for each agent. The maximum input length for the LLM during the reproduction process was limited to 10,000 tokens, while this limit was increased to 50,000 tokens for generating reports in the Issue Detector. We have implemented a conversation compression mechanism to handle cases where LLM-simulator interaction history exceeds the model's input token limits, based on related research [57]. When retrieving similar samples for the target review, the number of candidate samples $k$ was set to 10. For semantic-based retrieval, we employed the *sentence_transformers* library [11] with the pre-trained *paraphrase-MiniLM-L6-v2* [52] model to generate dense vector embeddings for textual data and assess similarity based on vector proximity. Our Android experimental environment was configured using the Genymotion [6] emulator running Android version 12.1. The framework interfaces with the emulator to implement control and operations via uiautomator2 [12], a Python library for Android UI automation, and the ADB [1].

## 5 Results

In this section, we present the experimental results by answering the following two research questions (RQs). For each RQ, we describe its motivation, outline our approach, and then discuss the result.

#### RQ1: How effective and efficient is RevPerf at reproducing performance issues from reviews compared with baselines?

**Motivation.** The goal of this RQ is to evaluate **RevPerf**'s performance along two key dimensions: its effectiveness in successfully reproducing real-world performance issues from user reviews, and its efficiency, measured by the time required for each successful reproduction attempt.

**Approach.** To assess the performance of **RevPerf**, we tested it on the manually reproduced dataset we have constructed in Section 4.1. Specifically, we apply **RevPerf** to read the original review, execute the potential steps to trigger the performance issue, and detect the issue features in the emulator. Following Huang et al. [27], we report the success rate, which is defined as the percentage of issues successfully reproduced within a predefined time frame (one hour). For efficiency, we measure the average processing time from review input to successful report generation for cases where issues are successfully reproduced. After analyzing the quantitative results, we also conducted a qualitative analysis to examine our model's performance in both successful and failed cases.

Table 1. Reproduction results on the manually reproducible performance-issue dataset. Human denotes the total time spent on manual reproduction. **AG** and **RB** denote the baseline approaches **AdbGPT** and **ReBL**, respectively. **-RA, -RA(VF), -RD, -LA, and -UI** denote **RevPerf** variants with the corresponding component removed. × indicates a failed reproduction attempt. * indicates a statistically significant difference from **RevPerf** (McNemar's exact test with Holm–Bonferroni correction, $\alpha = 0.05$).

| | RQ1 | | | | RQ2 | | | | | | RQ1 | | | | RQ2 | | | | |
|---|---|---|---|---|---|---|---|---|---|---|---|---|---|---|---|---|---|---|---|
| **Issue** | **RevPerf** | **Human** | **AG*** | **RB*** | **-RA** | **-RA(VF)*** | **-RD** | **-LA*** | **-UI** | **Issue** | **RevPerf** | **Human** | **AG*** | **RB*** | **-RA** | **-RA(VF)*** | **-RD** | **-LA*** | **-UI** |
| Aegis#1 | 163 | 102 | × | 134 | 135 | × | 149 | × | 259 | Kiwix#4 | 245 | 40 | × | 71 | 108 | × | 80 | 79 | 93 |
| Aegis#2 | × | 139 | × | × | × | × | × | × | × | Kiwix#5 | 69 | 86 | × | 36 | 62 | 94 | 95 | × | 174 |
| Aegis#3 | 206 | 84 | × | × | × | 231 | 224 | × | 263 | Kiwix#6 | 211 | 90 | × | × | 172 | 216 | 117 | 144 | 148 |
| AnkiDroid#1 | 89 | 65 | × | 51 | 71 | × | 94 | 72 | × | Kiwix#7 | × | 76 | × | × | × | × | × | × | × |
| AnkiDroid#10 | 97 | 36 | 42 | 86 | 80 | × | 134 | 94 | 112 | Markor#1 | 182 | 49 | × | 188 | 258 | 324 | × | 140 | 186 |
| AnkiDroid#2 | 151 | 83 | × | 152 | 150 | 277 | 158 | 168 | 234 | Markor#2 | 180 | 123 | × | 228 | 156 | × | 447 | × | 242 |
| AnkiDroid#3 | 224 | 64 | × | 113 | 170 | × | 143 | 200 | 192 | Markor#3 | 103 | 28 | 36 | × | 150 | 161 | 197 | 80 | 109 |
| AnkiDroid#4 | 269 | 131 | × | × | 171 | 172 | 116 | 263 | 217 | Markor#4 | 76 | 35 | × | × | 63 | 82 | 67 | 61 | 132 |
| AnkiDroid#5 | 124 | 41 | × | 92 | 225 | 180 | 109 | × | 134 | Markor#5 | 102 | 75 | × | 60 | 95 | 182 | 120 | 107 | 233 |
| AnkiDroid#6 | 209 | 129 | × | 181 | 110 | 145 | 166 | 139 | 129 | Markor#6 | × | 140 | × | × | × | × | × | × | × |
| AnkiDroid#7 | 219 | 81 | × | × | 192 | 390 | 229 | × | 220 | QKSMS#1 | 190 | 217 | × | 157 | 109 | 318 | 158 | 274 | 173 |
| AnkiDroid#8 | 147 | 109 | × | × | 149 | × | 155 | 443 | 263 | QKSMS#2 | × | 110 | × | × | × | × | × | × | × |
| AnkiDroid#9 | 251 | 103 | × | × | 258 | 274 | 246 | 202 | 233 | QKSMS#3 | 67 | 40 | 40 | 90 | 46 | 119 | 51 | 69 | 87 |
| AntennaPod#1 | 145 | 102 | × | × | 113 | 226 | 131 | 118 | 129 | QKSMS#4 | × | 55 | × | × | × | × | × | × | × |
| AntennaPod#2 | × | 439 | × | × | × | × | × | × | × | QKSMS#5 | × | 34 | × | × | × | × | × | × | × |
| AntennaPod#3 | 171 | 63 | × | 157 | 100 | 152 | 263 | 140 | 208 | QKSMS#6 | 98 | 104 | × | 133 | 104 | 100 | 227 | × | 196 |
| AntennaPod#4 | 125 | 74 | × | × | 96 | 153 | 115 | × | 131 | QKSMS#7 | 120 | 64 | × | 94 | 99 | 126 | 133 | 182 | 161 |
| AntennaPod#5 | 128 | 50 | × | × | 174 | 136 | 214 | × | × | QKSMS#8 | × | 53 | × | × | × | × | × | × | × |
| AntennaPod#6 | 106 | 100 | × | × | 222 | × | 299 | × | 120 | RandomNamePicker#1 | 144 | 65 | × | 85 | × | 187 | × | × | 171 |
| AntennaPod#7 | 175 | 68 | × | × | 98 | 167 | 133 | × | 147 | Simplenote#1 | × | 104 | × | × | × | × | × | × | × |
| AntennaPod#8 | 133 | 36 | × | 54 | 75 | 118 | 123 | 81 | 74 | Simplenote#2 | × | 71 | × | × | × | × | × | × | × |
| ChessClock#1 | 124 | 38 | × | 106 | × | × | 196 | 287 | 112 | Simplenote#3 | × | 97 | × | × | × | × | × | × | × |
| Kiss#1 | 133 | 45 | 124 | 162 | 180 | 148 | 102 | 71 | 129 | Simplenote#4 | × | 98 | × | × | × | × | × | × | × |
| Kiss#2 | 99 | 34 | 34 | × | 82 | 85 | 88 | 59 | 94 | Simplenote#5 | × | 127 | × | × | × | × | × | × | × |
| Kiss#3 | 136 | 42 | × | 193 | 136 | 116 | 144 | 146 | 115 | Simplenote#6 | × | 87 | × | × | × | × | × | × | × |
| Kiwix#1 | 113 | 95 | × | 143 | × | 162 | 122 | 176 | × | Simplenote#7 | 68 | 86 | × | × | × | 137 | 155 | 97 | 273 |
| Kiwix#2 | × | 92 | × | × | × | × | × | × | × | UserLAnd#1 | 172 | 193 | × | 121 | 153 | × | × | 123 | 241 |
| Kiwix#3 | 101 | 69 | × | 30 | × | 80 | 63 | 57 | × | | | | | | | | | | |

**Baselines.** As there are no existing baseline models that have similar functions to reproduce and detect performance issues, we adopt two recently proposed general LLM-based crash reproduction frameworks, AdbGPT [21] and ReBL [57], as our baselines. For a fair comparison, we evaluate the baselines under the same detection pipeline by attaching our unified performance-issue detector (log, UI, and resource-usage checks) to AdbGPT and ReBL, which is invoked after they complete the predicted reproduction steps. This setup also demonstrates the modularity of our detection component: it can be readily integrated into existing LLM-based bug report reproduction frameworks to provide performance-issue detection, indicating the extensibility of our approach.

**Results. In terms of effectiveness, RevPerf achieves a 72.73 % success rate in reproducing performance issues, outperforming the best baseline ReBL by 27.28 percentage points. The average reproduction time is 146.66 seconds.** The detailed results are shown in Table 1. The evaluation dataset spans all three performance issue categories, covering diverse real-world performance symptoms. **RevPerf** successfully reproduced 40 out of 55 performance issues. This result highlights the ability of our framework to capture key information from app reviews and detect their manifestation by monitoring logs, system-resource usage history, and GUI changes on the screen. In comparison, AdbGPT reproduced only 5 out of 55 issues (9.1 %) with an average time of 55.23 seconds, while ReBL reproduced 25 issues (45.45 %) with an average time of 116.72 seconds. The effectiveness improvements of **RevPerf** over both baselines are statistically significant under McNemar's exact test [49] with Holm–Bonferroni correction [26] (both corrected $p < 0.001$). AdbGPT struggled in most cases because its reproduction capability relies heavily on the completeness and quality of the input review text, whereas app reviews are typically incomplete and ambiguous. Although ReBL achieved a higher success rate than AdbGPT, it still fell short of **RevPerf**, primarily because it misses actions required for background and cross-app behaviors (e.g., lock/unlock, switching via the App Drawer, and resuming from the background). Moreover,

ReBL does not leverage information from related reviews to enrich the execution context, which further limits its ability to reproduce performance issues reported in reviews. To better understand where **RevPerf** falls short, we next conduct a detailed analysis of the failure cases and summarize the key reasons why reproduction did not succeed.

**Analysis of Failed Reproductions.** **RevPerf** failed to reproduce 15 performance issues from reviews. Among these failed cases, 12 were caused by limitations in UI exploration, while the remaining three required cross-version differential analysis, which is beyond the current design of **RevPerf**. This result suggests that UI exploration, rather than issue detection, accounts for most of the observed reproduction failures.

| |
|---|
| **Kiwix#2** |
| Please add an item in the settings that would disable 'instant search when entering text'. Instant search leads to application freezes. |
| **QKSMS#2** |
| App was great until my phone **updated** it yesterday automatically. Now when I type enough to go to a new line, it doesn't automatically scroll down until I pause typing for a second, so until I manually stop, I'm typing blind. … |
| **AntennaPod#2** |
| Great open source app, works on Android Auto. I used it years ago but was a little buggy before. Moved to Google podcasts then back to this one recently, and was able to import my whole subscription list which was good. One thing I notice now is the Episodes refresh feature is a little slow when you have 40 or more subscriptions, they possibly need to address that lag. Also once in a while it will replay the same 10seconds of audio but not often. Really has been improved over the years! |

Fig. 8. Case Studies of **RevPerf** Reproduction Failures.

**First, 12 failed cases are caused by limitations in the UI exploration component, which fails to drive RevPerf into the buggy state through realistic user interactions.** Of these, 11 require rapid, consecutive input sequences that cannot be faithfully reproduced by the current action space. For example, Kiwix#2 in Fig. 8 reports a freezing problem triggered by rapidly typing words into the search box using the virtual keyboard, and QKSMS#2 presents a similar input-lag issue. Although these issues can be reproduced manually, existing frameworks rely on functions such as `set_text` or `adb shell input text` to inject a sequence of characters in a single call, rather than generating fine-grained Input Method Editor (IME) events as a real user would. The remaining case is Aegis#2, where the framework needs to select images from a gallery app whose interface uses custom Canvas-based rendering. The image thumbnails are not exposed as separate accessibility nodes [4], so they are absent from the UI hierarchy obtained through `AccessibilityService` APIs that **RevPerf** relies on. In future work, we plan to extend the action space with a human-like IME typing module to support fine-grained interaction timing, and to incorporate visual UI understanding to handle dynamically rendered components that are invisible to accessibility-based UI hierarchies.

**Second, three performance issues can only be effectively detected through the comparison between different app versions.** An example is illustrated in Fig. 8. For performance issue AntennaPod#2, the user reported that a new update decreased the subscription refreshing speed. During manual reproduction, testers could detect this issue by comparing the time taken to update the same list across different app versions. However, automated reproduction of this entire process is challenging as it significantly increases operational complexity by doubling the required operations, introducing software version management through installation and uninstallation processes, and requiring the capture of corresponding key logs from different locations within extensive log data. We plan to introduce multi-simulator control mechanisms to reproduce the same operations across different software versions, enabling more complex comparative analysis.

**Runtime and API Cost.** We evaluate the efficiency of **RevPerf** in terms of reproduction time and LLM inference cost, as reported in Tables 1 and 2. The measurement covers the entire reproduction process, from reading the review to generating the final report. In Table 1, `Human` denotes the average time spent by the two authors on manual reproduction, which is 88.41 seconds. **RevPerf** takes an average of 146.66 seconds to complete the process, reflecting the overhead introduced by its three LLM-assisted stages. AdbGPT and ReBL are faster, with average reproduction times of 55.23 and 116.72 seconds, respectively. This is partly because they successfully reproduce fewer complex performance issues. As shown in Table 2, Feedback-Driven Execution accounts for the largest share of runtime (89.98 s, 61.35 %), while Issue Detection dominates both token usage (82.23 %) and API cost ($0.0392). Overall, **RevPerf** costs $0.0488 on average to complete one successful reproduction-and-detection run. These results indicate that **RevPerf** introduces additional runtime due to its LLM-assisted stages, but maintains a low API cost per run.

Table 2. Average runtime, LLM token usage, and API cost of **RevPerf** across reproduction stages. Results are averaged over the 40 successfully reproduced cases, with API costs estimated from Google Gemini API pricing. Token Share is computed as the percentage of the total tokens across all three stages.

| Stage | Time (s) | Tokens (In / Out) | Token Share (%) | Cost ($) |
|---|---|---|---|---|
| Review Augmentation | 18.36 | 2,194.70 / 705.70 | 0.98 | 0.0012 |
| Feedback-Driven Execution | 89.98 | 48,611.45 / 871.80 | 16.79 | 0.0084 |
| Issue Detection | 38.32 | 239,825.35 / 2,578.05 | 82.23 | 0.0392 |
| **Total** | **146.66** | **290,631.50 / 4,155.55** | **100.00** | **0.0488** |

### RQ2: How does each design component of RevPerf contribute to its overall effectiveness?

**Motivation.** In this RQ, we aim to evaluate the impact of each component on the overall performance of **RevPerf**. Given that **RevPerf** integrates several modules different from crash reproduction frameworks [21, 27, 57], we plan to remove or replace the components, including the ReviewAggregator, LogAnalyzer, UIInspector, and ResourceDiagnoser.

**Approach.** The ablation study is conducted by systematically disabling or simplifying key components of **RevPerf** and measuring the change in reproduction success. Specifically, we consider the following variants: **RevPerf**_RA, **RevPerf**_RA(VF), **RevPerf**_LA, **RevPerf**_RD, and **RevPerf**_UI. Each variant is applied to the same dataset of 55 performance issues evaluated in RQ1. Among these variants, **RevPerf**_RA removes the ReviewAggregator and uses the original review as the sole information source for reproduction, while **RevPerf**_RA(VF) disables the version-based filtering in the ReviewAggregator so that reviews from other versions of the same app can also be selected as candidate reviews. **RevPerf**_LA and **RevPerf**_RD remove the corresponding analysis modules and revise the related prompts accordingly. Finally, **RevPerf**_UI prevents the agent from using UI-state changes to determine whether a performance issue has been triggered, requiring it instead to rely on log messages or resource-usage changes as evidence.

**Result. Each component contributes to the overall performance of RevPerf.** The experiment results show that removing any single component degrades the overall performance of **RevPerf**. Table 3 compares the ablation results with the original **RevPerf** in terms of effectiveness and efficiency. Removing the ReviewAggregator component resulted in a trade-off between effectiveness and efficiency due to the lack of the first stage. The success rate dropped from 72.73 % (40/55) to 61.82 % (34/55) (15.00 % relative decrease), while the average reproduction time decreased from 146.66 s to 134.28 s (8.45 % faster). When the version-based filtering is disabled in the ReviewAggregator, the results become even worse than removing the whole module. Candidate reviews from other versions may be unrelated to the original review, resulting in noisy context for the reproduction. Furthermore, the longer context required more frequent LLM calls for history compression. This leads to a drop in the success rate from 72.73 % (40/55) to 54.55 % (30/55), which is a 25.00 % relative decrease, and the average reproduction time increases from 146.66 s to 175.21 s, which is 19.46 % longer. Removing LogAnalyzer reduced the success rate by 30.00 % and decreased the average reproduction time by 0.79 %. Removing ResourceDiagnoser reduced the success rate by 7.50 % and increased the average reproduction time by 6.19 %. Compared to the full **RevPerf**, the **RevPerf**_UI variant resulted in a 10.00 % lower success rate and took 16.18 % longer to complete. Statistically, removing LogAnalyzer and disabling version-based filtering lead to statistically significant degradations under McNemar's exact test with Holm–Bonferroni correction (both corrected $p < 0.01$). The remaining three variants also reduce the success rate, but do not reach statistical significance after correction, with only 3–6 discordant pairs in these comparisons. To illustrate the effectiveness of **RevPerf**, we conduct an analysis based on the success cases.

Table 3. Ablation Results in RQ2. The arrows in the Effectiveness column indicate the relative change in success rate compared with **RevPerf** (↓ lower, ↑ higher), and the arrows in the Efficiency column indicate the relative change in average reproduction time compared with **RevPerf** (↓ faster, ↑ slower). * indicates a statistically significant difference from **RevPerf** (McNemar's exact test with Holm–Bonferroni correction, $\alpha = 0.05$).

| Method | Effectiveness | Efficiency |
|---|---|---|
| **RevPerf** | 72.73 % | 146.66 s |
| **RevPerf**_RA | 61.82 % (↓ 15.00 %) | 134.28 s (↓ 8.45 %) |
| **RevPerf**_RA(VF)* | 54.55 % (↓ 25.00 %) | 175.21 s (↑ 19.46 %) |
| **RevPerf**_LA* | 50.91 % (↓ 30.00 %) | 145.51 s (↓ 0.79 %) |
| **RevPerf**_RD | 67.27 % (↓ 7.50 %) | 155.75 s (↑ 6.19 %) |
| **RevPerf**_UI | 65.45 % (↓ 10.00 %) | 170.40 s (↑ 16.18 %) |

**Analysis on Successful Detection.** First, **RevPerf** can identify app settings and system settings mentioned in reviews, and proactively check and correct inconsistencies between the mentioned and current settings. For example, Aegis#3 states: "It's almost perfect but please honor my Android settings with reduced animations, or offer a setting to disable them." These animations cannot be disabled within the app, resulting in user discomfort. During the Feedback-Driven Execution stage, **RevPerf** exits the app and navigates to the system settings to disable the animation, thereby providing the prerequisite for successful detection. With this adjustment, **RevPerf** successfully reproduced the animation-related performance issue described in the review.

Second, ReviewAggregator helps LLMs acquire richer context by retrieving other relevant reviews from the same app version. As an example, AntennaPod#1 reports: "During playback pause play button stops working after screen off and on. Must go back to Queue list and press play from there." The original review reports that the issue arises after the screen state changes, but lacks a reproducible action sequence and a clear description of the resulting app behavior. In this case, ReviewAggregator retrieves another review stating: "Whenever I try to use the audio controls to pause or play, the entire app stops responding. It does not crash however. When I close the app from the app list, it still continues to play in the background and I have no way of stopping the audio unless I force quit the app." This additional review provides a more explicit action sequence and clarifies the resulting behavior (e.g., UI unresponsiveness and background playback). With the augmented context, the LLM can generate a more concrete and actionable reproduction procedure for the following execution phase.

Moreover, our framework provides detection robustness through multiple analyzers. Performance issues can be detected by multiple analyzers in specific cases. For example, while the memory leak issue in QKSMS#1 (cf. Fig. 3) is successfully detected by monitoring memory usage, LogAnalyzer can also identify logs reporting "A resource failed to call close." Such findings demonstrate that our framework possesses complementary detection capabilities and does not rely entirely on a single component.

Table 4. Distribution and reproduction results of the 55 performance issues by category. Category totals are non-mutually exclusive, as some issues span multiple categories.

| Issue Category | Total | Success | Failure | Success Rate |
|---|---|---|---|---|
| Slow Interaction and Delayed Response | 36 | 22 | 14 | 61.11 % |
| App Freezing and Unresponsiveness | 28 | 23 | 5 | 82.14 % |
| Excessive Resource Consumption | 6 | 6 | 0 | 100.00 % |
| Overall | 55 | 40 | 15 | 72.73 % |

## 6 Discussion

In this section, we discuss failure cases encountered during manual reproduction, which we perform to construct a reproducible dataset of performance-issue reviews. We also summarize key findings from our experiments.

### 6.1 Failure Cases Study

In our collected dataset, 91.07 % (561/616) of the issues still cannot be successfully reproduced by the authors due to data requirements, even though they are confirmed by developers. We attribute the failures to two reasons as follows.

The first reason is the insufficient data volume required to trigger performance issues. For example, in the messaging application QKSMS, a user reported that a conversation containing more than 18,000 messages resulted in laggy and slow performance. Manually populating such extensive content is impractical, consequently leading to our failure in reproducing similar performance issues that depend on large-scale data volumes.

The second reason is the heterogeneity of data types necessary for reproducing performance issues. During attempts to reproduce WordPress performance problems, numerous users reported lag and freezing behaviors linked to the article "block" feature. Since blocks represent core compositional elements supporting diverse content types and styling variations, we faced significant challenges in determining the precise block configurations required to replicate the reported performance degradations.

### 6.2 Issue Type Distribution of the Evaluation Dataset

We further examine the composition of the 55 performance issues used in our evaluation. Table 4 presents the distribution of issues across categories along with the corresponding reproduction results. Since 13 issues belong to multiple categories, these category counts are not mutually exclusive. As shown in the table, the final dataset is not perfectly balanced across categories. This aligns with prior findings that slowness and unresponsiveness issues are reported more often than resource-consumption ones [38]. Issues involving **App Freezing and Unresponsiveness** and **Excessive Resource Consumption** achieve higher reproduction success rates than those involving **Slow Interaction and Delayed Response**. Specifically, **RevPerf** reproduces 23 out of 28 **App Freezing and Unresponsiveness** issues and all 6 **Excessive Resource Consumption** issues, while successfully reproducing 22 out of 36 **Slow Interaction and Delayed Response** issues. Overall, **RevPerf** achieves a reproduction rate of 72.73 %. Although the sample size for **Excessive Resource Consumption** is small, its high success rate may be partly explained by the presence of explicit technical cues in some reviews, such as "memory leak," which provide clearer signals for reproduction. We also observe that **App Freezing and Unresponsiveness** is reproduced more successfully than **Slow Interaction and Delayed Response**. A likely reason is that the former often requires less complex action sequences to trigger, which is consistent with our failed-case analysis in RQ1.

### 6.3 Insights from the Reproduction Study

**Conceptual Mismatch Between User Reports and Failure Modes.** User reviews are often noisy and symptom-oriented, and their wording may not match developer-defined failure modes. We illustrate this mismatch with the frequently used term "freeze": it is an umbrella term that can indicate multiple underlying behaviors. Our manual investigation reveals that "freeze" reports may correspond to several distinct situations:

- **True unresponsiveness (ANR-like behavior):** the app becomes genuinely non-responsive and appears frozen.
- **Long-running work on the UI without feedback:** a heavy computation is executed while the UI provides no progress indicator, causing users to suspect a dead loop or hang after waiting for a long time.
- **Lifecycle-driven perceived freeze:** a heavy task is tightly coupled to a UI component. When the app is backgrounded, UI updates may be paused, causing the app to appear frozen even if the background task continues and potentially pausing or stalling the task itself.
- **Unreliable background task execution:** tasks lack of reliable background task management (e.g., retry/backoff or continuation under poor network and resource pressure), leading to intermittent execution and user-perceived freezing.

These failure semantics become evident only when we execute and observe the app behavior during manual reproduction. Consequently, keyword-based matching is insufficient. Since such ambiguity cannot be reliably resolved from review text alone, static analysis provides limited diagnostic guidance, which further highlights the importance of automated methods.

**Limitations of Accessibility-Based UI Exploration.** Our failure analysis also reveals a limitation of accessibility-based UI exploration. Some app interfaces render internal components through custom drawing rather than exposing them as separate accessibility nodes. In such cases, components that are visible to users may not appear in the accessibility hierarchy, making them difficult for the agent to identify and interact with. This limitation suggests that combining accessibility-based exploration with vision-based UI understanding could better support dynamically rendered interfaces.

### 6.4 Implications for Practical Deployment

To construct our dataset, we collected 95,067 raw reviews from Google Play as the initial review pool. Through multiple preprocessing and filtering steps, we constructed a manually validated dataset of 55 reproducible performance issues, corresponding to a reproduction rate of 8.83 % relative to the filtered subset. In comparison, prior work using GitHub issues as input achieved a reproduction rate of 12.71 % [64]. We therefore view our result as a reasonable initial yield for this underexplored task, indicating that app reviews can serve as a viable starting point for surfacing user-perceived performance issues. At the same time, the relatively low yield suggests that practical deployment should rely on a selectively filtered review pipeline, and our results should be interpreted as initial evidence of feasibility on this filtered subset rather than as a generalizability claim over the broader population of app-store reviews.

## 7 Related Work

**Automated Bug Reproduction.** Prior work in automated reproduction research has focused on reproducing crashes and functional issues, with notable approaches including ReCDroid and ReCDroid+ [63, 64], ReproBot [62], MACA [41], DroidScope [28], and Yakusu [20]. These methods are limited by the quality of available information resources and the complexity of target applications.

To address these limitations, recent works have improved automated bug reproduction by leveraging LLM capabilities, from transforming bug reports into structured entities [21] to incorporating feedback mechanisms for enhanced accuracy [57], and ultimately enabling reproduction from minimal single-sentence descriptions [27]. These approaches predominantly rely on structured bug reports or GitHub issues. While such reports often contain richer technical details, they may not fully capture user-perceived performance problems arising in diverse real-world devices and usage contexts. To leverage more diverse types of information sources, RepRev [34] employed NLP techniques and exploration algorithms to extract crash-related entities from app reviews, achieving 69.84 % reproduction success.

**Android Performance Issue Analysis.** Several studies have analyzed Android performance issues from different perspectives. Liu et al. [45] conducted an empirical study on 70 real-world performance bugs from popular Android applications to characterize their properties and identify common bug patterns. They validated their findings by building a static analysis tool that successfully detected 126 new performance issues. Liao et al. [38] conducted an empirical study employing app reviews to investigate the root causes of mobile app performance issues and summarized existing datasets and approaches for analyzing performance issues. Kumari et al. [32] qualitatively analyzed 385 Stack Overflow Q&A posts to formulate two distinct taxonomies that categorize performance issues and their underlying causes in Android applications. While many studies have analyzed the characteristics of performance issues from different perspectives, most existing detection tools for general use still rely on static analysis [35, 46, 50] and dynamic analysis [17, 29]. There are limited studies that explore deep learning approaches for Android performance issue detection. Jabbarvand et al. [30] employed deep learning models to observe runtime features and locate energy anomalies. Recently, GUIWatcher [43] has been applied to detect performance issues such as GUI lags and long image loading times by analyzing key frames extracted from video recordings of app executions. However, these methods target specific apps and cannot be used for general detection across other apps.

## 8 Threats to Validity

**Internal Validity.** Several factors may affect the reliability of our experimental results. First, the inherent randomness of LLMs can introduce output variability, which we mitigated through multiple validation checks and structured prompt engineering. Second, the noisy nature of user reviews can also impact the accuracy of our semantic search for relevant information. To mitigate this threat, we use a secondary fine-grained filtering stage that also utilizes an LLM for more precise rule-based filtering. Third, the fixed retrieval budget $k = 10$ may also affect internal validity. Different $k$ values may change the amount of retrieved context and introduce different levels of noise. We set $k = 10$ as a practical choice to balance context coverage, prompt length, and inference cost rather than tuning it on the evaluation set. Since the retrieved reviews are further filtered by the LLM-based relevance filter, $k$ only defines the upper bound of the candidate pool. Future work will examine the sensitivity of the results to different $k$ values. Fourth, the accuracy of data labeling poses a potential threat to internal validity. Since we manually constructed the dataset and performed multi-class classification, the reliability of our annotation process may affect the experiment result. To mitigate this issue, we conducted initial test labeling and created a detailed guideline for the boundaries and overlapping cases. In the construction of the manually reproduced review dataset, one author attempted to reproduce 616 reviews in the simulator. To address potential misinterpretation of ambiguous review descriptions that could lead to false negatives, failed reproductions were independently validated by a second author. Finally, the adaptation of baselines may also affect internal validity, as AdbGPT [21] and ReBL [57] were originally designed for crash reproduction rather than performance issue reproduction. To mitigate this threat, we equipped both baselines with the same performance issue detector and aligned their detector prompts with those of **RevPerf**.

**External Validity.** External validity is primarily concerned with dataset representativeness, the influence of the experimental environment, the design scope of **RevPerf**, and the noisy, unstructured nature of app reviews. To enhance the generalizability of our findings, our dataset was constructed by referencing application lists from prior automated reproduction studies [21, 27, 57] and by limiting the number of reviews included from each application to ensure a more balanced selection. Nevertheless, the reproduction of performance issues is often highly sensitive to specific environmental configurations and data, meaning that results might vary across different settings despite efforts to standardize the testing environment. To obtain more complete and specific descriptions of the environment, we introduced a review augmentation phase. This phase utilizes semantics-based retrieval and LLM-based filtering to gather additional information from thematically similar review samples, thereby enabling a more detailed characterization of performance issues and their corresponding experimental environments. Beyond environmental variation, generalizability is also bounded by the design scope of **RevPerf**, which supports post-execution diagnosis rather than real-time performance monitoring or online intervention. Its lightweight tracing mechanism preserves transient runtime evidence for subsequent LLM-based analysis, but diagnosis reports are generated only after reproduction. Moreover, the relatively low yield of reproducible cases suggests that the current approach is better suited to selectively filtered reviews than to fully automatic large-scale deployment over raw app-store reviews. Another threat to external validity stems from the noisy and unstructured nature of app reviews, which differ substantially from more structured reporting channels such as GitHub issues. Consequently, our findings are most directly applicable to review-driven performance issue reproduction and may not fully generalize to these alternative reporting contexts.

## 9 Conclusion

Inspired by recent advancements in research on automated crash reproduction, this paper proposes **RevPerf**, a novel framework that leverages app reviews to obtain detailed descriptions of performance issues, automatically reproduce them, and flexibly detect their triggering with multiple methods. Using a curated dataset of 55 reproducible performance issues derived from 95,067 reviews across 29 apps, our framework **RevPerf** successfully reproduces 40 of these 55 issues, with an average reproduction time of 146.66 seconds. This work represents an initial step toward leveraging app reviews for automated performance issue reproduction, and demonstrates the potential to assist mobile application developers in identifying and addressing performance issues more efficiently. In future work, we plan to expand **RevPerf**'s applicability by integrating a wider array of detection methods and enriching it with more extensive background knowledge. Future efforts will also focus on incorporating more automatically adjustable environmental parameters to simulate more realistic and diverse experimental environments.

## 10 Data Availability

To support the Open Science policy, we have made our dataset, including the APKs and reproduction data, the source code of **RevPerf**, and the related documentation publicly available in a repository [37].

## Acknowledgments

This work was supported by the National Natural Science Foundation of China (Grant No. 62502409).